# Impact of V substitution on the physical properties of Ni-Zn-Co ferrites: structural, magnetic, dielectric and electrical properties

M. D. Hossain[1], A. T. M. K. Jamil[1], M. R. Hasan[2], M. A. Ali[3], I. N. Esha[4], M. A. Hakim[5], M. N. I. Khan[2*]

[1]Department of Physics, Dhaka University of Engineering and Technology (DUET), Gazipur, Bangladesh
[2]Materials Science Division, Atomic Energy Center, Dhaka 1000, Bangladesh
[3]Department of Physics, Chittagong University of Engineering and Technology (CUET), Chattogram 4349, Bangladesh
[4]Department of Physics, University of Dhaka, Dhaka 1000, Bangladesh
[5]Department of Glass and Ceramic Engineering, Bangladesh University of Engineering and Technology (BUET), Dhaka 1000, Bangladesh

*Corresponding author´s e-mail address: ni_khan77@yahoo.com (M.N.I. Khan)

**Abstract**

We have investigated the Vanadium- (V) substituted Ni-Zn-Co ferrites where the samples were prepared using solid-state reaction technique. The impact of $V^{5+}$ substitution on the structural, magnetic, dielectric and electrical properties of Ni-Zn-Co ferrites has been studied. XRD analysis confirmed the formation of a single-phase cubic spinel structure. The lattice constants have been calculated both theoretically and experimentally along with other structural parameters such as bulk density, X-ray density and porosity. The FESEM images are taken to study the surface morphology. FTIR measurement is also performed which confirms spinel structure formation. The saturation magnetization ($M_s$), coercive field ($H_c$) and Bohr magneton ($\mu_B$) were calculated from the obtained M-H loops. The temperature dependent permeability is studied to obtain the Curie temperature. Frequency and composition dependence of permeability was also analyzed. Dielectric behavior and ac resistivity are also subjected to investigate the frequency dependency. An inverse relationship was observed between the composition dependence of dielectric constant and ac resistivity. The obtained results such as the electrical resistivity, dielectric constants and magnetic properties suggest the appropriateness of the studied ferrites in microwave device applications.

# Introduction

Ferrites, influential ceramic materials are composed by combining, firing and blending a huge portion of $Fe_2O_3$ (iron (III) oxide) with a small amount of at least one metallic element, for example, manganese, zinc, nickel, barium, cobalt, etc. [1]. Ferrites mainly used in three sectors of electronics: power applications, low-level applications and Electro-Magnetic Interference (EMI) suppression. Therefore, a constant challenge is being posted to improve their characteristics for ever-increasing demands in home communication appliance, computer, electrical and other technical fields [2]. The interests in this oxide emerge from their versatile applicability from relatively high to microwave frequency region. Among various ferrites, soft ferrites being highly resistive at high frequencies are used extensively in electronic applications such as transformer cores, inductors, antenna rod, microwave devices, computer memory chip, magnetic recording media, etc. after their first commercial introduction [3]. Mostly these electrical and magnetic properties depend on the method of preparation, preparative parameters, preparative conditions, particle size, nature of dopants etc. [4]. Hence, many researchers bestowed their time and efforts on various ferrites with many dopants to enhance its electric and magnetic characteristics.

The important electronic properties make the spinel Ni-Zn-Co ferrites prominent [5]. A good combination of its magnetic and electric properties as well as it low-cost aspect makes them potential candidates for application in high-frequency purpose [6]. These ferrites have high saturation magnetization, low eddy current loss, high resistivity, high permeability, high Curie temperature [7-10]. It is technologically sound due to its prospective use in targeted microwave devices, sensors, catalysis, magnetic recording applications and drug delivery systems [11-15]. Owing to the mentioned interest many researchers have paid their attention to the Ni-Zn-Co ferrites. Mallapur *et al.* have investigated the structural and electrical properties of the spinel ferrite system of Ni-Co-Zn [16]. Electric properties of nanocrystalline Ni-Co–Zn ferrites have been reported by Ghodake et al [17]. Spectral studies such as room temperature Mossbauer, X-ray and infrared IR spectra of Ni-Co–Zn ferrites have been carried out by Amer *et al.* [18]. Knyazev *et al.* carried out a study on the structural and magnetic properties of Ni-Co–Zn ferrites [19]. Besides, Mohit *et al.* [20], Stergiou *et al.* [21], Ghodake *et al.* [22], Hassan *et al.* [23], Mattei *et al.* [24], Chen *et al.* [25], etc. have also investigated the Ni-Co–Zn ferrites. Moreover, reports on the different ions substituted-Ni-Zn-Co ferrites are also available e.g., Ren *et al.* [26]

have performed a study on $La^{3+}$ substituted Ni-Zn-Co ferrites. Saini *et al.* [27] have investigated In substituted Ni-Co-Zn ferrites. Y and La Substituted Ni-Co-Zn ferrites have been investigated by Stergiou *et al.* [28], Gd and La-doped Ni-Zn-Co ferrites are studied by Zhou *et al.* [29] BaO-doped Ni-Zn-Co magneto-dielectric ferrites have been studied by Zheng et al [30]. We have also studied the Gd substituted Ni-Zn-Co ferrites [31].

The dependency of the physical properties of ferrites on the distribution of cation over tetrahedral (*A*) and octahedral (*B*) sites open the way of changing their properties by introducing different ions into these two sites [32-34]. Therefore, alteration of physical properties by the incorporation of a small amount of V into Ni-Zn-Co is also expected. Reports on the addition or substitution of vanadium into different ferrites systems are available [35, 36]. Korkmaz *et al.* [37] noted a decline of saturation magnetization due to V substitution in $NiFe_{2-x}V_xO_4$ (x ≤ 0.3) nanoparticles system. Magnetic properties of V substituted Li-Zn-Ti ferrites have been investigated by Maisnam *et al.* [38]. They reported decrease in Curie temperature ($T_c$) and saturation magnetization with an increase in V contents. Slimani *et al.* [39] have investigated how magnetic and optical properties $NiFe_{2-x}V_xO_4$ (0.0 ≤ x ≤ 0.3) NPs are influenced by varying calcination temperature. Heiba *et al.* [40] studied the V substituted Co ferrites. M. Kaiser also reported the effect of V substitution on the magnetic and dielectric properties of Ni-Zn-Cu ferrites [41]. The decrease of saturation magnetization owing to the presence of V with the low melting temperature which causes the formation of the liquid phase, accelerating the grain growth process at low sintering temperature [42, 43]. Based on the available report, a considerable alteration of properties of Ni-Zn-Co ferrites is expected through V substitution. Therefore, the objective of the current study is to investigate the influence of V substitution on the electrical and magnetic properties of Ni-Zn-Co ferrites.

**EXPERIMENTAL DETAILS**

Conventional double sintering method was used to prepare $V^{5+}$ substituted Ni-Zn-Co ferrite [$Ni_{0.7}Zn_{0.2}Co_{0.1}Fe_{2-x}V_xO_4$ (0 ≤ *x* ≤ 0.12)]. The following operation are used to prepare the desired samples the details of which can be found elsewhere [31]. However, the sample preparation procedure is shown in Fig. 1.

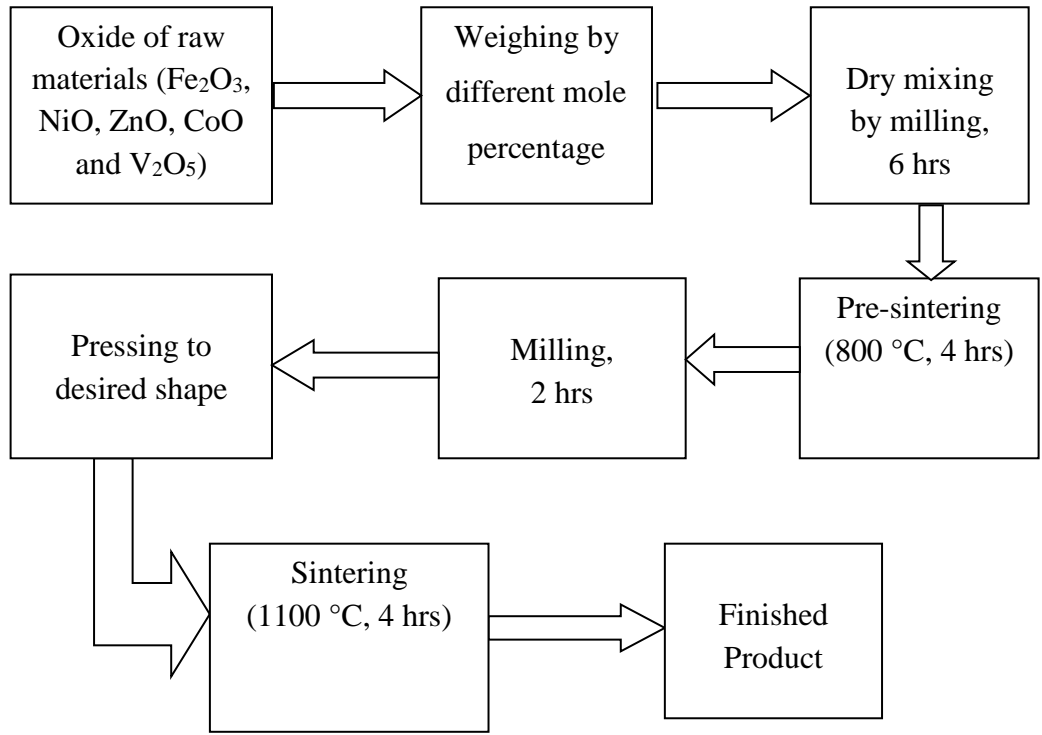

**Fig. 1:** The sample preparation procedure of $Ni_{0.7}Zn_{0.2}Co_{0.1}Fe_{2-x}V_xO_4 (0 \leq x \leq 0.12)$ ferrites.

The synthesized samples have been studied by X-ray diffraction (XRD) using Philips X'pert PRO X-ray diffractometer (PW3040) with Cu-K$_\alpha$ radiation (λ=1.5405 Å), Field Effect Scanning Electron Micrographs (FESEM) (JEOL JSM-7600F), Fourier transform infrared spectroscopy (FTIR) (PerkinElmer FT-IR Spectrometer). A Wayne Kerr precision impedance analyzer (6500B) [Frequency range: 1kHz–120 MHz, drive voltage of 0.5 V] was also used to study the dielectric and permeability properties. The magnetic properties were obtained by a physical properties measurement system (PPMS), Quantum Design Dyna Cool at ambient conditions.

**Results and discussion**

**Structural properties**

Fig. 2 illustrates the X-ray diffraction (XRD) patterns of $Ni_{0.7}Zn_{0.2}Co_{0.1}Fe_{2-x}V_xO_4$ ($0 \leq x \leq 0.12$) ferrites sintered at 1100 °C. The diffraction peaks at different planes (111), (220), (311), (222), (400), (422), (511) and (440) confirmed the cubic spinel structures (JCPDS #08-0234) of the synthesized samples [44]. The diffraction peaks become broaden with the increase of the concentration of V doped which indicates the decrease of grain size [Table 1] with the increase of V contents. It is occurred because of the smaller ionic radius of $V^{5+}$ ion than the ionic radius of $Fe^{3+}$ ion.

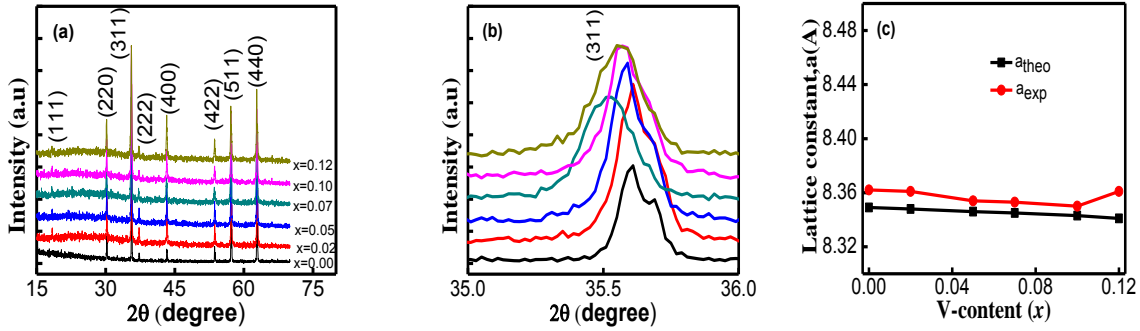

**Fig.2: (a)** X-ray diffraction patterns **(b)** Maximum pick broaden and **(c)** Lattice constant($a$) dependence of V-content of $Ni_{0.7}Zn_{0.2}Co_{0.1}Fe_{2-x}V_xO_4 (0 \leq x \leq 0.12)$ ferrites.

The experimental lattice constant ($a$) for all the samples are calculated using the formula [45]:

$$a_{exp} = d_{hkl}\sqrt{h^2 + k^2 + l^2} \qquad (1)$$

where, h, k and l are the Miller indices of the crystal planes. The calculated values are given in Table 1. The obtained value of $a_{exp}$ for $x = 0.00$ is 8.362 Å and the reported value is 8.3719 Å [26]. Our obtained value is lower than the reported value, might be due to different synthesis condition. The change of lattice constant ($a$) with V content is shown in Fig.2(c). At first $a_{exp}$ decreases with the increase of V contents and then increases for $x = 0.12$. The trend in the variation lattice constant is observed to have a good agreement with earlier report [46] of V-substituted Ni–Zn ferrites. The decrease of the lattice constant is owing to the differences of ionic radii (Shannon 6-coordination numbers) of $V^{5+}$ions (0.59Å) and $Fe^{3+}$ions (0.645Å) [47,48]. The lattice constant of studied samples $Ni_{0.7}Zn_{0.2}Co_{0.1}Fe_{2-x}V_xO_4$ ($0 \leq x \leq 0.12$) has also been calculated to compare with the experimental one. The lattice constant can be calculated theoretically using the following equation [2]:

$$a_{th} = \frac{8}{3\sqrt{3}}\left[(r_A + R_0) + \sqrt{3}(r_B + R_0)\right] \qquad (2)$$

where, $R_0$, $r_A$, $r_B$ are the ionic radii of oxygen (1.32 Å) [49], $A$- and $B$-sites, respectively. The $r_A$ and $r_B$ can be calculated by the following equations:

$$r_A = C_{AZn}r(Zn^{2+}) + C_{AFe'}r(Fe^{3+}) \text{ and } r_B = \frac{1}{2}[C_{BNi}r(Ni^{2+}) + C_{BCo}r(Co^{2+}) + C_{BFe}r(Fe^{3+}) + C_{BV}r(V^{5+})].$$

[50, 51].

The calculated X-ray density, bulk density and porosity for the studied samples $Ni_{0.7}Zn_{0.2}Co_{0.1}Fe_{2-x}V_xO_4$ ($0 \leq x \leq 0.12$) were represented in Table 1. The equations used can be found elsewhere [31]. A decreasing trend of X-ray density and the bulk density with increasing V-content till $x = 0.12$ is noted while an inverse relation is followed by the porosity. This trend can be explained on the basis of atomic mass of $V^{5+}$ (50.94 amu) and $F^{3+}$ (55.84 amu). Here, V with lower atomic mass is substituted for Fe ions with comparatively higher atomic mass. Whereas, for $x = 0.12$ the increase in density was observed due to the increase of lattice constant. Moreover, the $\rho_b$ is found to be lower than the $\rho_x$, because the pores are considered in the calculations of bulk density which are absent in the x-ray density calculations [52].

Fig. 3 demonstrates FESEM images from where the changes in the microstructure owing to the V substitution can be observed. In addition, the average grain size for considered compositions has been estimated [shown in Table 1] by linear intercepting method and a decreasing trend with V contents is also observed except $x = 0.12$ which is similar to the variation of density and porosity with V contents. The changes in the average grains size, density and porosity can be understood from the FESEM images. The change in average grains size is also related to the difference in between ionic radii of $V^{5+}$ (0.59 Å) ions and $Fe^{3+}$ ions (0.645 Å).

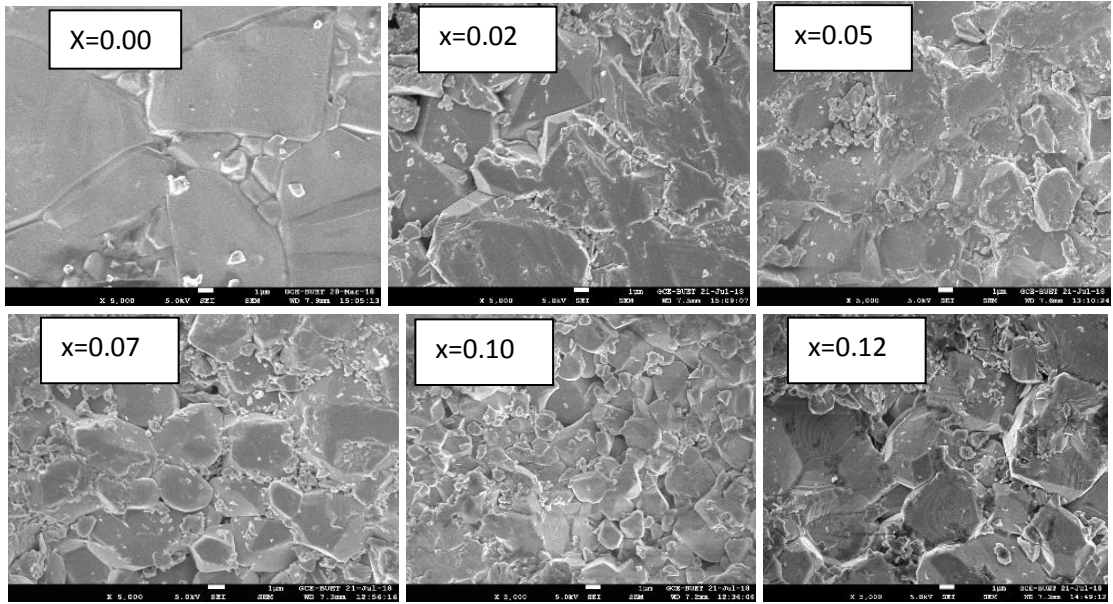

**Fig. 3:** The FESEM images of $Ni_{0.7}Zn_{0.2}Co_{0.1}Fe_{2-x}V_xO_4$ ferrites.

**Table 1:** Cation distribution (tetrahedral A-site and octahedral B-site), Ionic radii ($r_A$ for A-site and $r_B$ for B-site), Lattice constant ($a_{th}$ for theoretical and $a_{exp}$ for experimental value), X-ray density ($\rho_x$), Bulk density ($\rho_b$) and Porosity (P) of $Ni_{0.7}Zn_{0.2}Co_{0.1}Fe_{2-x}V_xO_4$ ($0 \leq x \leq 0.12$) ferrites.

| V content (x) | Cation Distribution | | Ionic radii | | Lattice constant | | X-ray density | Bulk density | Grain Size D | Porosity | Bond Length (Å) | | Hoping Length (Å) | |
|---|---|---|---|---|---|---|---|---|---|---|---|---|---|---|
| | A-site | B-site | $r_A$ (Å) | $r_B$ (Å) | $a_{th}$ (Å) | $a_{exp}$ (Å) | $\rho_x$ (g/cm³) | $\rho_b$ (g/cm³) | (μm) | P (%) | A-site | B-site | A-site | B-site |
| 0.00 | $(Zn^{2+}_{0.2}Fe^{3+}_{0.8})_A$ | $[Ni^{2+}_{0.7}Co^{2+}_{0.1}Fe^{3+}_{1.2}]_B O^{2-}_4$ | 0.664 | 0.666 | 8.349 | 8.362 | 5.35 | 5.05 | 6.87 | 5.71 | 1.810 | 2.091 | 3.621 | 2.956 |
| 0.02 | $(Zn^{2+}_{0.2}Fe^{3+}_{0.8})_A$ | $[Ni^{2+}_{0.7}Co^{2+}_{0.1}Fe^{3+}_{1.18}V^{5+}_{0.02}]_B O^{2-}_4$ | 0.664 | 0.665 | 8.348 | 8.361 | 5.51 | 4.26 | 4.43 | 22.68 | 1.810 | 2.090 | 3.620 | 2.956 |
| 0.05 | $(Zn^{2+}_{0.2}Fe^{3+}_{0.8})_A$ | $[Ni^{2+}_{0.7}Co^{2+}_{0.1}Fe^{3+}_{1.15}V^{5+}_{0.05}]_B O^{2-}_4$ | 0.664 | 0.664 | 8.346 | 8.354 | 5.49 | 4.19 | 3.73 | 23.68 | 1.808 | 2.089 | 3.617 | 2.954 |
| 0.07 | $(Zn^{2+}_{0.2}Fe^{3+}_{0.8})_A$ | $[Ni^{2+}_{0.7}Co^{2+}_{0.1}Fe^{3+}_{1.13}V^{5+}_{0.07}]_B O^{2-}_4$ | 0.664 | 0.664 | 8.345 | 8.353 | 5.48 | 4.11 | 3.70 | 25.0 | 1.808 | 2.088 | 3.617 | 2.953 |
| 0.10 | $(Zn^{2+}_{0.2}Fe^{3+}_{0.8})_A$ | $[Ni^{2+}_{0.7}Co^{2+}_{0.1}Fe^{3+}_{1.1}V^{5+}_{0.1}]_B O^{2-}_4$ | 0.664 | 0.663 | 8.343 | 8.350 | 5.46 | 4.08 | 2.58 | 25.27 | 1.807 | 2.088 | 3.616 | 2.952 |
| 0.12 | $(Zn^{2+}_{0.2}Fe^{3+}_{0.8})_A$ | $[Ni^{2+}_{0.7}Co^{2+}_{0.1}Fe^{3+}_{1.08}V^{5+}_{0.12}]_B O^{2-}_4$ | 0.664 | 0.662 | 8.341 | 8.361 | 5.48 | 4.43 | 3.98 | 19.16 | 1.810 | 2.090 | 3.620 | 2.956 |

## FTIR analysis

FTIR Spectroscopy is an important technique to study the completion of the solid-state reaction and inspect the presence of deformation in the spinel ferrites due to substitutions of ions [53, 54]. In case of ferrites, there are two characteristic peaks in the FTIR spectra. The first peak at low frequency is associated M-O (Metal-Oxide) stretching vibrations at B-sites while the peak at higher frequency side results from M-O stretching vibrations at A-sites [55]. Fig. 4 displays the FTIR spectra of V substituted Ni-Zn-Co ferrite compositions in which the peaks are observed at expected positions, confirmed the completion of solid state reaction [56]. It has two major absorption bands in the range of 365 to 800 cm$^{-1}$. The band's position for the studied compositions is tabulated in Table 2. The bands at higher frequency ($\upsilon_1$ = 590 cm$^{-1}$) is for stretching vibrations of the M-O clusters at the tetrahedral site. Lower frequency bands ($\upsilon_2$ = 365cm$^{-1}$) are assigned to the stretching mode of the M-O bond in the octahedral sites [57].

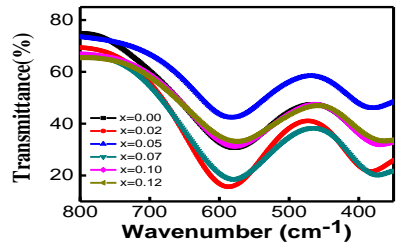

**Fig. 4:** FTIR spectra of Ni$_{0.7}$Zn$_{0.2}$Co$_{0.1}$Fe$_{2-x}$V$_x$O$_4$ (0 ≤ $x$ ≤ 0.12) ferrites.

**Table 2:** The vibrational frequencies of two prominent IR bands corresponding to tetrahedral and octahedral sites of Ni$_{0.7}$Zn$_{0.2}$Co$_{0.1}$Fe$_{2-x}$V$_x$O$_4$ (0 ≤ $x$ ≤ 0.12) ferrites.

| V-content ($x$) | $\upsilon_1$(cm$^{-1}$) | $\upsilon_2$(cm$^{-1}$) |
|---|---|---|
| 0.00 | 578.49 | 372.19 |
| 0.02 | 590.09 | 379.98 |
| 0.05 | 583.69 | 381.18 |
| 0.07 | 583.69 | 370.99 |
| 0.10 | 579.69 | 373.58 |
| 0.12 | 574.70 | 364.59 |

## Magnetic properties

Fig. 5 shows the magnetic hysteresis (M-H) loops of Ni$_{0.7}$Zn$_{0.2}$Co$_{0.1}$Fe$_{2-x}$V$_x$O$_4$ (0 ≤ $x$ ≤ 0.12) ferrites. The parameters such as saturation magnetization ($M_s$), coercivity ($H_c$), magnetic remanence ($M_r$), etc were obtained from the M-H loops and listed in Table 3. The narrow hysteresis loops assure the soft magnetic nature of all the samples [58]. In the present study for un-doped Ni-Zn-Co ferrites, the value of saturation magnetization is 71.6 emu/g. It is observed to decrease due to V substitution because of having non-magnetic nature. The $V^{5+}$ ions have a liking to occupy $B$-sites. Although, the magnetic moment of ferrites is dominated by A-B interaction but contribution from A-A interaction and B-B interaction is noticeable. The net magnetic moment is given by the equation: $M = M_B - M_A$, where $M_B$ represents the total magnetic moment at B-sites and the $M_A$ indicates the total magnetic moment at A-sites. The total magnetic moment of B-sites is lowered by replacing Fe ions by V substitution and it caused the reduction of magnetic moment of ferrites system. Therefore, the lowering of saturation magnetization due to V substitution is reasonable. However, the $M_s$ is observed to increase for $x = 0.12$, might be owing to the increased grain size at $x = 0.12$ because of the proportional relationship between magnetization and grains size [59]. The values of $M_r$ also follow the similar trend with V contents. Another important parameter, known as Bohr magneton ($\mu_B$) is also calculated using the relation: $\mu_B = \frac{M' M_S^r}{5585}$, $M'$ is the molecular weight. The calculated values of $\mu_B$ also follow the trend of $M_s$ with V contents.

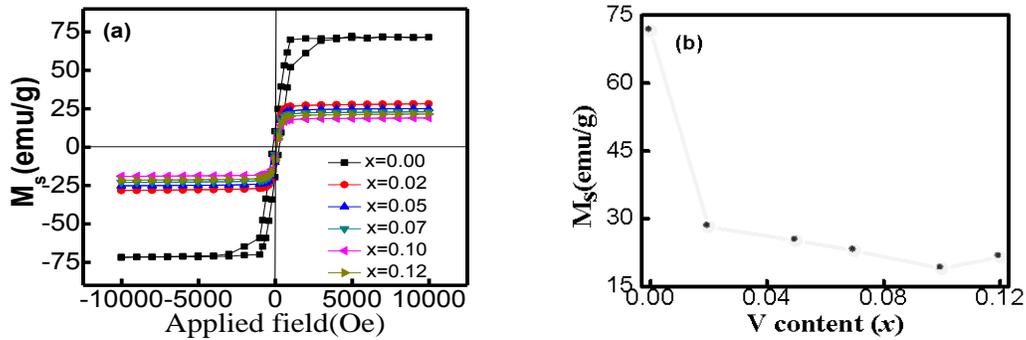

**Fig.5:** The variation of saturation magnetization dependence on **(a)** applied field and **(b)** V content($x$) for $Ni_{0.7}Zn_{0.2}Co_{0.1}Fe_{2-x}V_xO_4$ ($0 \leq x \leq 0.12$) ferrites.

**Table 3 :** Saturation magnetization ($M_s$), Magnetic remanence ($M_r$), Resonance frequency ($f_r$) coercivity ($H_c$), Magnetic moment ($\mu_B$), Curie Tem. ($T_c$), Real permeability ($\mu'$) and Relative quality factor (RQF) for $Ni_{0.7}Zn_{0.2}Co_{0.1}Fe_{2-x}V_xO_4$ ($0 \leq x \leq 0.12$) ferrites.

| V content ($x$) | Saturation Magnetization $M_s$(emu/g) | $M_r$ (emu/g) | $f_r$ (MHz) | Coercivity $H_c$ (Oe) | $n_B$ in $\mu_B$ Theo | $n_B$ in $\mu_B$ Exp. | Curie Tem. $T_c$ (°C) | $\mu'$ | RQFx$10^3$ |
|---|---|---|---|---|---|---|---|---|---|
| 0.00 | 71.6 | 22.08 | 60.97 | 277.2 | 3.70 | 3.21 | 478 | 253 | 5.11 |
| 0.02 | 28.24 | 9.42 | 56.92 | 20.39 | 3.66 | 2.78 | 476 | 237 | 27.5 |

| | | | | | | | | |
|---|---|---|---|---|---|---|---|---|
| 0.05 | 25.16 | 7.96 | 59.00 | 27.36 | 3.60 | 2.36 | 474 | 226 | 67.6 |
| 0.07 | 22.95 | 7.46 | 59.00 | 29.5 | 3.56 | 1.99 | 458 | 213 | 164.3 |
| 0.10 | 18.98 | 6.22 | 49.91 | 20.26 | 3.50 | 1.86 | 436 | 184 | 203.2 |
| 0.12 | 21.55 | 5.36 | 61.20 | 15.64 | 3.46 | 2.15 | 465 | 202 | 220.9 |

## Study of temperature dependent permeability: Curie temperature

The magnetic properties such as saturation magnetization, permeability etc. are very sensitive to temperature. One of the characteristic parameters, Curie temperature ($T_c$) can be obtained from the temperature dependent permeability. The permeability remains almost constant up to certain temperature and after increasing slightly by exhibiting a peak, known as Hopkinson's peak, it finally drops sharply to be zero. The temperature at which the permeability becomes zero is known as Curie temperature ($T_c$), the temperature at which completely disorderliness of atomic moments took place and the ferrimagnetic materials converted to paramagnetic. The temperature dependent initial permeability was measured and shown for different composition in Fig. 6(a). The mentioned features are also observed for our obtained data and the calculated Curie temperature is shown in Fig. 6 (b).

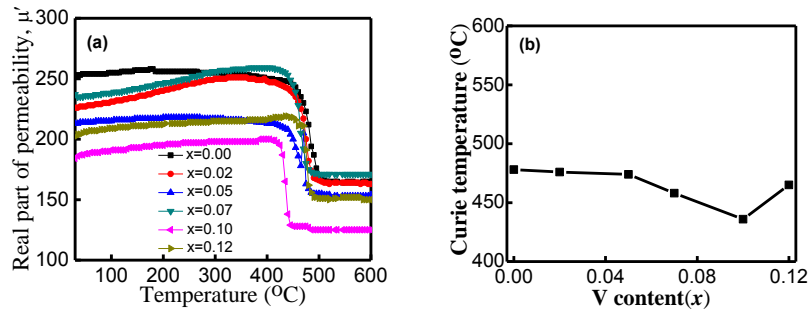

**Fig. 6:** Variation of (**a**) permeability with temperature and (**b**) V content (*x*) dependence curie temperature for $Ni_{0.7}Zn_{0.2}Co_{0.1}Fe_{2-x}V_xO_4 (0 \leq x \leq 0.12)$ ferrites.

The measured $T_c$ is noted to decline with the V substitution. The transition temperature attributed from the distribution of cation over *A*- and *B*-site and value of exchange coupling constant (*J*) [60]. The $J_{AB}$ attributed from interaction between ions on *A*- and *B*-sites is much stronger than the $J_{AA}$ and $J_{BB}$ attributed from interaction between ions of same sites (*A or B*). The replacement of $Fe^{3+}$ ions by $V^{5+}$ ions, decrease the magnetic ions in the B-sites causing a decrease in the strength of exchange coupling constant $J_{AB}$ and thus decrease the $T_c$ values.

## Frequency dependence of real permeability and Relative Quality Factor

Fig. 7 (a) shows the real part of permeability ($\mu'$) as a function of frequency. The toroid shaped samples were prepared for this characterization. The value of $\mu'$ remains almost constant up to ~ 30 MHz, a noticeable increase is noted at which the $\mu'$ became maximum and then fall sharply to certain low values. The steady value is important for many applications such as in transformer as a broadband pulse and in video recording system as read-write heads (wide band) [61].

In different circumstances, a significant peak is exhibited by $\mu''$ [figure is not shown] at the frequency where sharp decline of $\mu'$ is taken place. This phenomenon is termed as ferrimagnetic resonance [62] and the prepared compositions are found to follow the Snoek's limit [63].

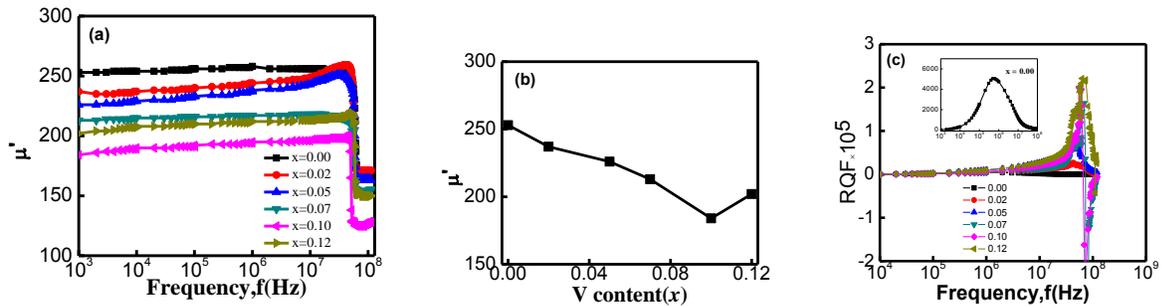

**Fig. 7:** (a) Frequency dependence of initial permeability ($\mu'$), (b) V-content ($x$) dependence of real permeability and (c) Frequency dependent relative quality factor (RQF) of $Ni_{0.7}Zn_{0.2}Co_{0.1}Fe_{2-x}V_xOs_4 (0 \leq x \leq 0.12)$ ferrites.

Fig. 7(b) illustrates the composition dependence of $\mu'$ which revealed that the initial permeability decreased gradually with V contents up to $x = 0.10$ and then increased slightly for $x = 0.12$. The permeability of ferrites depends on various factors like grain size, density, porosity, saturation magnetization, anisotropy, etc. [64]. A good correlation is observed among permeability, average grain size [Table-1] and saturation magnetization [Fig.5(b)]. The Globus-Duplex relation $\mu' \propto \dfrac{M_s^2 D}{\sqrt{K_1}}$, (where, $M_s$, $D$ and $K_1$ represent the saturation magnetization, average grain size and anisotropic constant, respectively) can be used to explore the variation of $\mu'$ with V contents. The above relation exhibits that the initial permeability is proportional to $M_s$ and $D$ where $M_s$ and $D$ are also found to decrease with V contents up to $x = 0.10$ and then increases for $x = 0.12$.

The relative quality factor (RQF) measures the performance of a material for use in filter. Fig. 7 (c) shows the usual behavior of RQF versus frequency for different V contents in which a prominent peak is observed for each composition [31, 65, 66]. The exhibition of peak is followed by very low value both in the low frequency and in the high frequency regions. The declination of RQF corresponds to the frequency (~30 MHz) wherein the $\mu''$ (loss component) started to increase sharply. Moreover, the value of RQF is observed to increase owing to V substitution with the maximum for $x = 0.12$. The values of RQF are given in Table 3.

## Dielectric behavior

A very common dielectric behavior of ferrites is that the $\varepsilon'$ and $\varepsilon''$ decrease with frequency but exhibit three different responses in three ranges of frequencies: (i) a sharp decrease with low frequency in $\varepsilon'$ and $\varepsilon''$ ($\varepsilon''$ is not shown), (ii) a slow decrease within mid frequency range in $\varepsilon'$ and $\varepsilon''$ and (iii) finally become almost frequency independent at high frequencies [31,65,67,68]. Comparatively slower decrease in $\varepsilon'$ and $\varepsilon''$ is observed in the mid-frequencies region where the orientational polarization is mainly responsible for dielectric constant. The dielectric constant becomes constant at the very high frequency region where the contribution comes from the atomic and electronic polarization [69]. The frequency dependence of dielectric constant ($\varepsilon'$) and dielectric loss factor ($tan\delta = \varepsilon''/\varepsilon'$) of $Ni_{0.7}Zn_{0.2}Co_{0.1}Fe_{2-x}V_xO_4$ ($0 \leq x \leq 0.12$) ferrites are shown in Fig. 8 (a and b). The dielectric properties can be explained by considering the studied ferrites as composed of two layers: grains of high conductivity which are separated by grain boundary of high resistivity [70-72]. The hopping mechanism is responsible of dielectric mechanism [73, 74]. The charges produced through electron hopping between $Fe^{3+}$ and $Fe^{2+}$ are responsible for electrical conduction in ferrites. The motion of these charges are supposed to be hindered at the grain boundaries owing to their activity in the low frequency region and accumulated at the interface causing space charge polarization. Although, the grains are active at the high frequency region but electron hopping cannot follow the frequency of applied external electric that causes a reduction of the charges produced by hopping between $Fe^{3+}$ and $Fe^{2+}$ [75]. Therefore, the decrease of $\varepsilon'$ and $\varepsilon''$ with increasing frequency is expected. From Fig. 8 (a), it is obvious that the dielectric constant is decreased due to V substitution which is explained later.

Fig. 8 (b) demonstrates the unusual behavior of dielectric loss (tanδ) as a function frequency. The curves *tanδ* vs frequency exhibit a maximum at a particular frequency but different with V concentration variation. This peak is observed to shift to the high frequency side owing to the increasing V substitution. The similar abnormal behavior (of tanδ) has also been reported for V substituted Ni-Zn-Cu ferrites [41], Sn substituted Ni-Zn ferrites [65] and Y substituted Mg-Zn [68]. This type of maximum in tanδ usually occurred when the jumping frequency of electron hopping becomes approximately equal to that of the externally applied ac electric field [76]

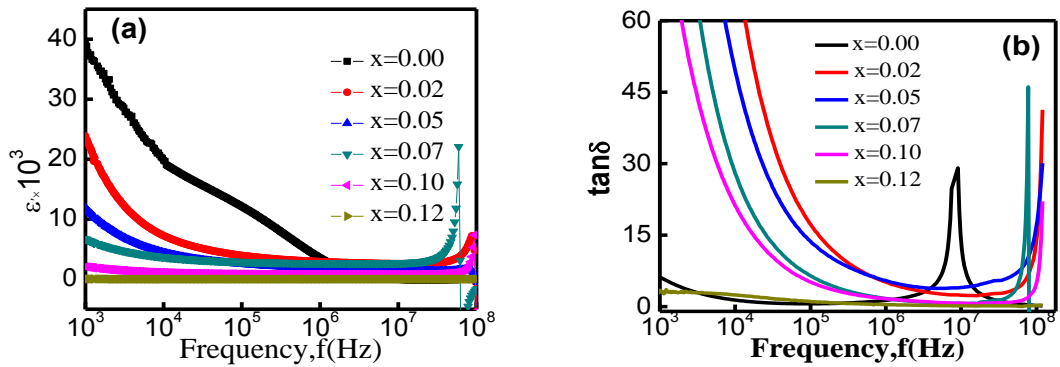

**Fig.8:** Frequency dependence of (**a**) dielectric constant and (**b**) dielectric loss of $Ni_{0.7}Zn_{0.2}Co_{0.1}Fe_{2-x}V_xO_4 (0 \leq x \leq 0.12)$ ferrites.

## Frequency dependence of AC resistivity

The frequency dependent ac resistivity ($\rho_{ac}$) is shown in Fig. 9(a) in the frequency range 1 kHz - 120 MHz. Again, the normal behavior in the frequency dependence of ac resistivity is observed for each sample. The resistivity is observed to decrease with increasing frequency and then becomes invariant at high frequency [65]. Fig. 9 (a) also demonstrates that the ac resistivity is noted to increase with V contents as shown in Fig. 9 (b). It is seen from Fig. 8 (a) and 9 that the dependence of dielectric constant and ac resistivity on V contents follows the opposite trend that is normal trends in ferrites. The dependence of dielectric constant or ac resistivity on V contents can be explained by assuming two mechanisms. Firstly: The V is substituted for $Fe^{3+}$ ions at the *B*-sites that lead to the decrease of electronic hopping between $Fe^{3+}$ and $Fe^{2+}$ occurs at B-sites due to reduced $Fe^{3+}$ ions. Thus, the decrement of electronic conductivity attributed from the electronic hopping mechanism is expected. Such type of reducing electronic conductivity is also reported by means of $Fe^{3+}$ ions replacement with ions that have a tendency to occupy the B-sites

[65]. Secondly: The decrement of grain sizes with V contents also contributed to the enhanced resistivity. The FESEM micrographs shows that the average grains size is observed to be decreased [Table 1] with V contents that leads to the increased in number of grain boundaries, the more grain boundaries results more insulating barriers in the way of charge carriers [69]. Consequently, the decrease of grain size also revealed the reduction of the conductive area. [77] Hence, the electrical resistivity is expected to be increased owing to V substitutions as shown in Fig. 9 (b).

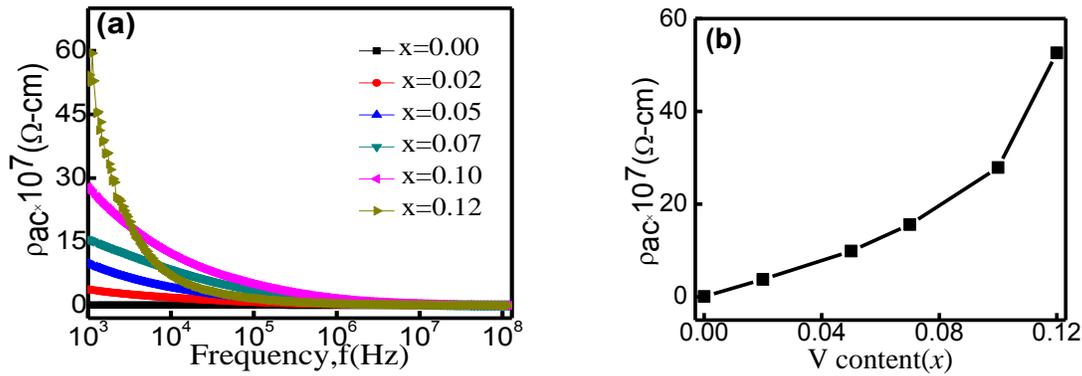

**Fig. 9:** (**a**) Frequency dependence and (**b**) V content($x$) dependence of ac resistivity of $Ni_{0.7}Zn_{0.2}Co_{0.1}Fe_{2-x}V_xO_4$ ($0 \leq x \leq 0.12$) ferrites.

## Conclusion

$Ni_{0.7}Zn_{0.2}Co_{0.1}Fe_{2-x}V_xO_4$ ($0 \leq x \leq 0.12$) ferrites have been synthesized by conventional ceramic technique. The decrease of lattice constant with V content is observed. A good correlation is observed among lattice constant, density and porosity for different V contents. Average grains size is found to decrease with V contents. The formation of spinel cubic ferrites is confirmed from the obtained values of vibrational frequency $\upsilon_1$ (in the range: 574 to 590 cm$^{-1}$) and $\upsilon_2$ (in the range: 364 to 382 cm$^{-1}$). The soft ferromagnetic nature is observed from hysteresis loops with low values of coercive field. The lowering of $M_s$ (from 71.6 emu/g to 18 – 28 emu/g) is due to non-magnetic V substitution. The Curie temperature obtained from temperature dependent permeability is observed to decrease owing to V substitution by means of decreasing the strength of exchange coupling constant. The study of the frequency dependent permeability exhibits a good correlation among the permeability, average grain size and saturation magnetization. The dielectric constant is found to decrease in V-substituted Ni-Zn-Co ferrites where the ac resistivity varies inversely with V contents. It is also expected that the results obtained in this paper will encourage the materials scientist to investigate the effect of V substitution on other ferrites system.